\documentclass{emulateapj}
\usepackage{amsmath}
\usepackage{graphicx}  
\usepackage{hyperref}
\usepackage{color}

\newcommand\unit[1]{\, {\rm #1}}
\newcommand\pasa{PASA}
\newcommand\nar{New A Rev.}
\newcommand\eqnbreak[1]{ \nonumber \\ &#1& }

\newcommand{\bi}{\begin{itemize}}
\newcommand{\ei}{\end{itemize}}
\newcommand{\be}{\begin{enumerate}}
\newcommand{\ee}{\end{enumerate}}

\newcommand{\gsim}{\gtrsim}
\newcommand{\expnt}[2]{\ensuremath{#1 \times 10^{#2}}}   
\newcommand{\Msun}{\ensuremath{M_{\odot}}}
\newcommand{\ergs}{\ensuremath{{\rm erg\,s}^{-1}}}

\newcommand{\peryr}{\ensuremath{{\rm yr}^{-1}}}

\newcommand{\er}{\ensuremath{\epsilon_{\rm radio}}}
\newcommand{\eedd}{\ensuremath{\epsilon_{\rm Edd}}}

\shorttitle{Detecting Merging SMBHs with Radio Surveys}
\shortauthors{Kaplan et al.}

\begin{document}

\title{Blindly Detecting Merging Supermassive Black Holes with Radio Surveys}
\author{D.~L.~Kaplan, R.~O'Shaughnessy}
\affil{Physics Dept. and Center for Gravitation and Cosmology,, U. of Wisconsin - Milwaukee, Milwaukee
  WI 53211}
\email{kaplan@uwm.edu, oshaughn@gravity.phys.uwm.edu}
\author{A. Sesana}
\affil{Albert Einstein Institute, Am Muhlenberg 1 D-14476 Golm,
  Germany; and Center for Gravitational Wave Physics, The Pennsylvania
  State University, University Park, PA 16802}
\email{alberto.sesana@aei.mpg.de}
\and
\author{M. Volonteri}
\affil{University of Michigan, Astronomy Department, Ann Arbor, MI,
  48109}
\email{martav@umich.edu}
\slugcomment{ApJL, in press}

\begin{abstract}
Supermassive black holes presumably grow through numerous mergers
throughout cosmic time.  During each merger, supermassive black hole
binaries are surrounded by a circumbinary accretion disk that imposes
a significant ($\sim 10^4\,$G for a binary of $10^8\,M_{\odot}$)
magnetic field.  The motion of the binary through that field will
convert the field energy to Poynting flux, with a luminosity $\sim
10^{43}\unit{erg\,s}^{-1}\,(B/10^4 G)^2(M/10^8 M_\odot)^2$, some of
which may emerge as synchrotron emission at frequencies near 1\,GHz
where current and planned wide-field radio surveys will operate.  We
find that the short timescales of many mergers will limit their
detectability with most planned blind surveys to $<1$ per year over
the whole sky, independent of the details of the emission process and
flux distribution.  Including an optimistic estimate for the radio
flux makes detection even less likely, with $<0.1$ mergers per year
over the whole sky.  However, wide-field radio instruments may be able
to localize systems identified in advance of merger by gravitational
waves.  Further, radio surveys may be able to detect the weaker
emission produced by the binary's motion as it is modulated by
spin-orbit precession and inspiral well in advance of merger.
\end{abstract}

\keywords{black hole physics---cosmology: observations---radio continuum: general---surveys}

\section{Introduction}
Based on relativistic simulations incorporating force-free
electromagnetic fields, \citet*{pll10} suggest that mergers of
supermassive black holes (SMBHs) which occur in the presence of an
accretion disk may have significant Poynting flux.  This Poynting flux
may be detectable as an electromagnetic (EM) counterpart to the
gravitational wave (GW) signature of the merger 
(other mechanisms  have been proposed as direct and indirect electromagnetic
signatures of merger;
see \citealt{schnittman11} and references therein).
These mergers will also produce GW signatures,
accessible to the \textit{Laser Interferometer Space Antenna}
(\textit{LISA}) (for BH masses $M$ in the range $\simeq
[10^3,10^7]M_\odot$) and for exceptionally low masses to ground-based
GW detectors ($M\lesssim 10^3 M_\odot$; see, e.g.,
\citealt{2009PhRvD..80l4026R}).
Whether measured via GW or EM, the measured merger history will strongly constrain our understanding of the formation and evolution of supermassive
black holes (\citealt{2011PhRvD..83d4036S}; \citealt*{2007MNRAS.377.1711S}).

Only recently have radio surveys moved beyond inhomogeneous archival
data sets to systematic examinations of the variable sky
\citep[e.g.,][]{lgw+08,cbk+11,ofb+11}, and the situation will continue
to improve.  Advances in receivers and digital processing make
instantaneous fields-of-view of $>10\,{\rm deg}^2$ possible at GHz
frequencies, enabling repeated surveys of wide areas of the sky.
These technologies are being implemented as part of Square Kilometer
Array pathfinders under construction \citep{jbb+07,bdbjf09}.

While all 
searches for compact object mergers have so far been  
negative \citep[e.g.,][]{aaa+10}, the improving performance of both
gravitational \citep[see][]{adligo} and electromagnetic surveys
increases the discovery potential for a wide range of events.
In this \textit{Letter} we consider the detectability of the merger
event with radio surveys centered near frequencies of 1\,GHz.  We show
that the flare itself is very unlikely to be detected in the current
generation of radio surveys, largely independent of the amount of EM
flux emitted.  However, prior to the flare there could be other
modulation present which may be detectable.  In what follows, we use a
flat $\Lambda$CDM cosmology with $\Omega_{\rm M}=0.27$ and $h=0.72$.

\section{Electromagnetic Counterparts of Merger Flares}

\citet{pll10} simulated the merger of two $10^8\,\Msun$ BHs.  They
found a flare of Poynting flux (with $L\simeq 4\times
10^{43}\unit{erg/s}$ over $\approx 5$ hours) that occurred at the same time as
the GW emission peaked.  They also found lower-level emission before the
flare ($L\simeq 10^{43}\ergs$).  \citet{nlp+10}  interpreted the
pre-merger secular emission as two steady jets powered by the motion
of each black hole through the background magnetic field, with
luminosity $\propto (v/c)^2 B^2 M^2$.  For unequal masses, we
physically expect the luminosity to be provided by the faster, smaller black
hole moving through the magnetic field.  Using the model of
\citet{nlp+10}, if the more massive black hole has mass $M$ and the
less massive has mass $qM$, we expect a luminosity $L\propto q^2M^2$.

The choice of magnetic field directly affects the electromagnetic
luminosities inferred from these simulations.  Conservatively,
\citet{pll10} chose a magnetic field that limited their jet luminosity
to a small fraction $L\sim 0.002 L_{\rm Edd}$ of the Eddington
luminosity at merger.\footnote{The limiting magnetic field required to
   reach this luminosity ($B\simeq 6\times 10^{4}\,{\rm G}
   (M/10^8\,M_\odot)^{-1/2}$) is
   substantially smaller than the magnetic field created by the
   magneto-rotational instability (MRI) at the inner edge of the
   circumbinary disk, which we estimate to be $\sim 10^6\,{\rm
     G}(\alpha M/10^8 M_\odot)^{-7/20}$
   \citep{2006PhRvL..97v1103P,2007MNRAS.375.1070B}. Rather than adopt
   this large circumbinary field, we implicitly absorb uncertainties into
   the ill-determined efficiency $\epsilon_{\rm Edd}$.}
We adopt the same assumption: a jet
luminosity limited to a small fraction $\eedd=0.002$ of the Eddington limit at
the merger event:
\begin{eqnarray}
L_{\rm flare} = \eedd L_{\rm Edd}
\end{eqnarray}
for $q=1$, while for other mass ratios we assume that $L\propto q^2$.

Given the expected range of magnetic fields
($B=\expnt{6}{4}\,(M/10^8\,M_\odot)^{-1/2}\,$G, with black hole masses
going from $10^{3}\,M_\odot$ to $10^{10}\,M_\odot$), electrons
advected with the flow might emit synchrotron radiation near 1 GHz, as
mentioned by \citet{pll10}.  Thus, the merger flare could be a distinctive
radio signature out to  cosmologically significant distances:
\begin{eqnarray}
\label{eq:Scalefac}
d_{L,\rm Edd} &\simeq& \sqrt{L/4\pi F_{\rm min}} \eqnbreak{\simeq} 14.2
\unit{Gpc} \sqrt{q^2(M/10^6
  M_\odot)\frac{\er(\eedd/0.002)}{(F_{\nu,{\rm min}}/\unit{mJy})(\nu/{\rm
      GHz})}}
\end{eqnarray}
(corresponding to $z\approx 2$)  where for simplicity we assume
$F_\nu\propto L/\nu$.   In this expression, rather than model the emission mechanism (i.e., spectrum, beaming) in
detail, we assume a fraction $\er$ of this energy is emitted isotropically in radio frequencies.  
Given the modest  Lorentz factor and magnetic field, beaming
is not likely to be too strong.  As much as possible in what follows,
we attempt to give results that are independent of $\er$.

Though the emission \emph{spectrum} is uncertain, the emission \emph{duration} is not: it scales with the total mass of
the system, 
as it depends on the orbital timescale near merger.  Based on \citet{pll10}, we estimate
the merger flare duration by
\[
\tau_{\rm flare}\approx
5\unit{hr}\left(\frac{M}{10^8\,M_{\odot}}\right).
\]
We adopt this estimate for all mass ratios, since the orbital (and
hence merger) timescale is determined by the more massive BH.

\subsection{Merger Rates}
To assess the visibility of merger flares, we  employ a
merger rate distribution that depends on black hole masses and redshift.
As each comparable-mass merger doubles the black hole mass, given the
masses and growth timescales over which they assemble, the
supermassive black hole merger rate must be both low, less than
$10^{-8}\unit{Mpc}^{-3}\unit{yr}^{-1}$, and strongly biased towards
low-mass mergers: only a few merger events occur per year on our past
light cone.

The assembly of SMBHs is reconstructed through  Monte-Carlo
merger simulations, following  the hierarchical structure
formation paradigm.  These models evolve the BH population
starting from BH ``seeds,'' through accretion episodes triggered by
galaxy mergers, and include the dynamical evolution of SMBH-SMBH
binaries.  The SMBH population is consistent with observational
constraints, e.g.,  the luminosity function of quasars at $1<z<6$, the
$M-\sigma$ relation and the BH mass density at $z=0$
\citep*{2003ApJ...582..559V,2008MNRAS.383.1079V,2010MNRAS.409.1022V}.
We adopt two of the  fiducial
merger distributions  used in \citet{abb+09}: models LE and SE, where S
versus L refers to the seed size -- large or small -- and E refers to
``efficient'' accretion; see
\citealt{2011PhRvD..83d4036S}.
These models are representative of  a range of plausible SMBH growth scenarios.   
As with uncertainties in $\er$, we attempt to make our conclusions
robust to specific merger assumptions.

\section{The Visibility of Merger Flares}
\begin{figure}
\plotone{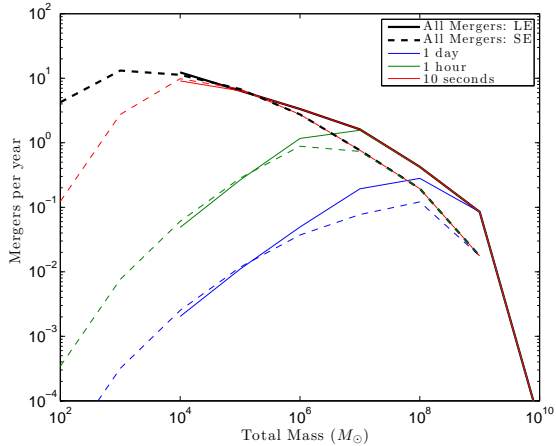}
\caption{Rate of mergers per year on our light cone, for a
  range of total system masses.  The thick curves include  all mergers for two models from  \citet{abb+09}, with solid
  lines based on the ``LE'' model and dashed lines based on the ``SE''
  model.  The thin curves reduce this number by $\text{max}(\tau_{\rm flare}/T,1)$ for  cadences $T=$ 1\,day (blue),
  1\,hour (green), and 10\,sec (red), assuming the flare event
  duration  $\tau_{\rm flare}\simeq 5\,{\rm h}(M/10^8\,M_\odot)$.  }
\label{fig:mergers}
\end{figure}

Figure \ref{fig:mergers} shows the total merger rate as
a function of BH mass, integrating over redshifts 0--10.  Only a few
mergers per year are expected, even from low-mass ($<10^6\,M_\odot$)
systems.  This rate is relevant to untriggered searches by all-sky
detectors such as GW observatories (LIGO, \textit{LISA}), which survey
the entire sky with roughly uniform sensitivity at high duty cycle.
For simplicity, in what follows we will provide quantitative results
primarily for the LE model; results from the two models are
comparable for the purposes of this discussion

For limited-aperture surveys, other factors limit the detectable fraction of events
(see the discussion in \citealt*{clm04}, for example).  
Ignoring any flux limits, two
effects are important. First, surveys only cover a fraction of the sky
$\Omega/4\pi$, with smaller coverage in each pointing.
For example, the  curvature of the Earth restricts
telescopes at temperate latitudes to $\Omega/4\pi\lesssim 80$\%; individual surveys will cover less.

Second, surveys return to the same area of the sky with a specific
cadence $T$.  A telescope which  surveys a single area continuously
(i.e., field-of-view $\Delta \Omega=\Omega$) 
has $T\simeq 0$.
More commonly $\Delta \Omega \ll \Omega$ and the telescope spends time
doing other tasks.  For instance, a survey might cover
$\Omega=10,000\,{\rm deg}^2$ with 333 pointings of $\Delta
\Omega=30\,{\rm deg}^2$, each lasting $30\unit{s}$.  The survey
finishes in $<3\unit{hr}$ (the smallest possible cadence).  If the
survey returns to each individual pointing 24\,hr later, the cadence
is $T=24\unit{hr}$.

With such a survey, the fraction of events that can be detected is the
fraction that happen to occur when observations are ongoing: ${\rm
  min}[\tau_{\rm flare}(1+z)/T,1]$, assuming $\tau_{\rm flare}$ is
much longer than both each pointing and any dispersive delay across
the bandpass (see \S~\ref{sec:GW}) and simplifying the flare emission
as either on or off \citep[e.g.,][]{cbk+11b}.  In
Figure~\ref{fig:mergers} we illustrate how this simple cadence cutoff
reduces the fraction of low-mass merger flares that could be found on
the past light cone of a survey with cadences $T=$ 1 day, 1 hour, and
10 seconds.  Though many low-mass mergers should occur, the short
durations of their merger flares makes them nearly impossible to
identify.

In the LE (SE) model, the total number of mergers on our
past light cone (summing over all redshifts, mass ratios, and masses)
is about 24 ({39}) per year.
A survey with $T=10\,$s cadence would recover most of the events, as
even the short low-mass events are sufficiently stretched by cosmology
that they would be visible for $M\gsim 10^4\,M_\odot$.  However, a
1\,h survey is only expected to see {3} ({2}) mergers per year out to
$z=10$ over the whole sky; surveys with finite area
will see correspondingly fewer.  Restricting the cadence to  1\,d
reduces the accessible number further, to {$0.5\,\peryr$}
({$0.2\,\peryr$}).

\subsection{Flux Distribution}
So far, we have only counted the number of mergers on the light cone
of our survey.
Using the predicted luminosities from \citet{pll10}, we provide in
Figure~\ref{fig:mergersflux} the cumulative rate of events greater
than a  flux threshold for a limiting cadence of 1\,s (to make
the numerous low-mass mergers visible).

Figure~\ref{fig:mergersflux} shows that merger flares are rare
events; the flux density corresponding to $>1\,\peryr$ over the
whole sky is only 0.13\,mJy (assuming efficiencies of $\er=1$ and
$\eedd=0.002$)\footnote{We 
  assume approximately isotropic emission.  If the
  emission is tightly beamed, single events will be detectable further
  away, but fewer events will be visible on our lightcone.  Since
   the number and timescale of events on our lightcone
   most limit potential surveys, strong beaming will reduce the numbers
  considered here.}.  The brightest events (tens of mJy) are extremely
rare and come from high-mass mergers, generally at moderately high
redshift.

In this figure, we have also limited the maximum possible duration of a merger flare 
to $10^6\,$s (roughly 12 days).  Longer events
are both rare -- excluding them changes little  -- and 
will be increasingly difficult to localize in time and separate from  systematic trends.

\section{Discussion \& Conclusions}
Because high-mass mergers are rare (though long-lasting) and low-mass mergers produce short and faint flares (though common), 
the rate of potentially detectable merger flares is small.
Even with optimistic choices for the efficiencies $\er$ and \eedd, we
expect {$<1$} merger per year with the
surveys to be conducted in the next decade, consistent with
zero detections to date.  Greater flux sensitivity will not increase
the detectable rate substantially,
 as the finite numbers and short timescales limit detectability.
Our results depend only weakly on 
the  assumed merger rate: while 
the low-mass and low-redshift merger rates are weakly constrained observationally, their merger flares will  rarely be visible. 

In the radio, ongoing and planned wide-field surveys have
instantaneous fields-of-view of $1-30\,{\rm deg}^2$
\citep[e.g.,][]{cba+10,jbb+07} at GHz frequencies (this increases to
several hundreds or even 1000\,deg$^2$ at a few hundred MHz).
Some have relatively frequent sampling, and cover the same area of the
sky on timescales from minutes to months, but generally only cover a
total of $<10^3\,{\rm deg}^2$.  Surveys that cover a wider area will
likely have a cadence of at least 1\,day.  None has the combination of
a very rapid cadence (ideally $<1\,$min) and very wide sky coverage
($>10^4\,{\rm deg}^2$) that are likely necessary to detect a flare
blindly, especially with a required sensitivity of $<0.01\,$mJy.
Since the instantaneous fields-of-view and cadences of planned optical surveys
are typically less than or comparable to those of radio surveys and
the cadence considerations are independent of wavelength, 
optical surveys will be unlikely to discover events like
these.  Only at X-ray and $\gamma$-ray energies would planned
instrumentation be well-suited to the timescales and rates of merger
flares, although the low fluxes ($\sim
8\times10^{-5}(M/10^8\,M_\odot)(E_{\rm photon}/10\,{\rm
  keV})^{-1}\,{\rm photon\,cm^{-2}\,s}^{-1}$ at a redshift of 0.1)
might require a next-generation mission to be detectable.

\subsection{GW Counterparts}
\label{sec:GW}
\begin{figure}
\plotone{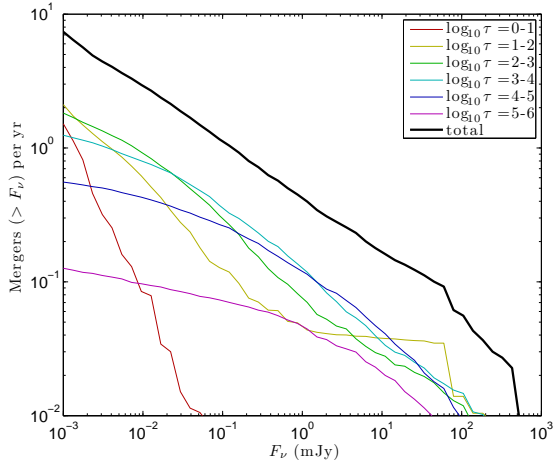}
\caption{Cumulative rate of mergers per year (based on the ``LE''
  model) brighter than a given flux density $F_{\nu}$ in mJy, for
  logarithmic bins of observed duration $\tau=\tau_{\rm flare}(1+z)$
  in seconds.  The thin lines are for each bin, while the thick line
  is for the total considering durations of $10^0$\,s to $10^6\,{\rm
    s}$.  We assumed timescales for the flare event that scale as
  $\tau_{\rm flare}\approx 5\,{\rm h}(M/10^8\,M_\odot)$, along with
  efficiency $\er=1.0$ and a frequency $\nu=1\,$GHz.  }
\label{fig:mergersflux}
\end{figure}

While Figure~\ref{fig:mergersflux} suggests that radio flares associated with mergers will be difficult to detect, 
next-generation surveys may reach
limits more amenable to detections.   We should consider how merger
flares could be identified as such and what physics may be learned
from them.

Unlike many proposed counterparts to SMBH mergers, this prompt
emission mechanism might allow coincident detection of electromagnetic
and GW signals from the same event, even though the
circumbinary disk is evacuated and no accretion onto the compact
objects takes place.  Spatial and temporal coincidence can confirm that a radio transient is indeed the signature
of a binary SMBH merger.  As reviewed in
\cite{whitepaper-CoordinatedScience}, coincident electromagnetic and
gravitational signals  provide an independent cosmological distance
ladder, if accessible at cosmological distances
\citep{2005ApJ...629...15H}.  
Additionally, nearby EM counterparts might be localized to
individual host galaxies, allowing study of galaxy-SMBH relations
\citep{schnittman11}.

In contrast with electromagnetic surveys, GW detectors have
roughly uniform all-sky sensitivity at all times and a signal that is
visible long before merger.  Using the GW signal as a
trigger, 
electromagnetic observations would be freed of the need to survey the whole sky continuously; followup observations
would be  
limited by  flux thresholds
alone.  Here, the large pointing uncertainties on
current-generation GW facilities (ideally $\sim
100\unit{deg}^2$ for a $<10^3 M_\odot$ SMBH merger with signal-to-noise of 8; e.g.,
\citealt{2009NJPh...11l3006F}) will make optical follow-up difficult
\citep[e.g.,][]{2009CQGra..26i4032H}, but are well suited to the
fields-of-view of instruments such as ASKAP \citep{jbb+07}.  Moreover,
the GW signal may allow identification of an impending merger well in
advance.  For prime \textit{LISA}-scale sources
($10^5-10^7 M_\odot$), 
the sky
location of an inspiralling binary can be located 
to within $10\unit{deg}^2$ hours to weeks before the merger
event (\citealt*{2008NewAR..51..884M}; \citealt{2007PhRvD..76b2003K}).  For the merger
trees discussed in this paper, this translates to several events per
year (slightly less than $1/3$ of all \textit{LISA}-detectable events)
that can be localized this precisely \citep{abb+09}.  For an
optimistic conversion of electromagnetic to radio energy, followup
pointings will identify all \textit{LISA} events with a flare only if
they reach a flux sensitivity $0.01\unit{mJy}(\eedd/0.002)\er$.  Less
sensitive followup observations will recover only a fraction of
events:
roughly $\simeq 17[1-0.4 (\log F_{\nu,{\rm min}}/\unit{mJy} -0.5)]$ events per year
for $F_{\nu,\rm min}\in 0.01-3\unit{mJy}$ and our fiducial efficiencies,
including all mass ratios.

For lower-mass mergers ($M\simeq 10^3 M_\odot$), ground-based GW
detectors will not identify  a sky location before the GW merger signal. 
 Nonetheless, if quickly processed, their sky localization can
still help target EM followup, as dispersion delays the EM signal.
Plasma dispersion will occur in the Milky Way, in the host galaxy of
the SMBH, and along the line-of-sight in the intergalactic medium.
For cosmological sources the total dispersion measure (DM, the
integral of the electron column density) may reach $>1000\,{\rm
  pc\,cm}^{-3}$ \citep{inoue04}.  This implies a time delay $\Delta
t=4.15\nu_{\rm GHz}^{-2}{\rm DM}\,{\rm ms}$, requiring rapid
localization and repointing.  Even for short flares with modest
dispersions, such a delay would be hard to detect, but it is possible
if the radio cadence is sufficiently short or the observing frequency
low.  It may also be possible to detect dispersion within the radio
data itself, by measuring the relative delays of different frequencies
across a bandpass of width $\Delta \nu$ ($\delta t=8\nu_{\rm
  GHz}^{-3}\Delta \nu_{\rm GHz} {\rm DM}\,{\rm ms}$).  This is more
difficult, since across a finite bandpass the relative delay is even
smaller, but is routinely done
\citep[e.g.,][]{lbm+07}.  In fact, for very low mass events $M\lesssim
10^4\,M_{\odot}({\rm DM}/1000\,{\rm pc\,cm}^{-3})(\nu/1\,{\rm
  GHz})^{-3}(\Delta \nu/300\,{\rm MHz})$ dispersive smearing will
exceed $\tau_{\rm flare}$, but this will not greatly change
Figure~\ref{fig:mergers}.  However, matching this ``internal'' delay
with that relative to GW observations could prove a powerful
confirmation of the nature of the event.

Second-generation ground-based GW detectors will be
sensitive to the lowest-mass mergers ($M\simeq 200-10^3\,M_\odot$) out
to a strongly mass- and orientation- dependent threshold $z\simeq
0.1-2$.
The associated EM flares will be short (dispersion-limited) and faint.
With the most optimistic efficiencies $\eedd,\er$, radio surveys would
have comparable reach to GW surveys at $F_\nu
\sim0.1\unit{mJy}$ (Eqn.~\ref{eq:Scalefac}); with less efficient
conversion or followup, fewer coincident events can be found.
Unfortunately, unlike SMBH mergers, observations do not directly
constrain such merging binaries.  The low-mass mergers to which
ground-based detectors are sensitive simply may not occur.  Even if
they do, both GW and radio observations are sensitive
to a minute fraction of the universe (not true for
third-generation GW detectors; \citealt{2009ApJ...698L.129S}).  That
said, if radio surveys can distinguish short ($<1\,$s) flares
in  targeted observations, ground-based GW detectors
working in concert with radio telescopes can rule out extremely
optimistic ($\gsim10^{-8}\unit{Mpc}^{-3}\unit{yr}^{-1}$) low-mass SMBH
merger rates and efficiencies.

\subsection{Non-Merger Events}
While the rate of potentially detectable mergers is small, 
other EM emission associated with binary SMBH inspiral could be detectable.
  EM emission from SMBHs is well known across a
range of wavelengths;  radio emission from active galactic nuclei
is common.   We  differentiate between generic AGN
emission and that associated with an orbiting pair of SMBH through the time domain.
AGN do vary intrinsically but mostly aperiodically;  detecting such periodic behavior in a
radio light curve would be a strong indication of an inspiralling SMBH
pair \citep[e.g.,][]{komossa06}; we defer additional methods for
confirmation to a forthcoming paper (O'Shaughnessy et al.\ 2011, in
prep.).  A number of binary AGN are known or suspected
\citep[e.g.,][]{komossa06,rtz+06,ssb+10,bs10}.  Most of these have
evidence from a resolved pair of bright spots or a double set of
emission lines, but they all probe systems far from actual merger
\citep{bs10}.  We consider what might happen as the systems approach
merger.

Variability will happen over a range of timescales.  First, even
before the merger the EM flux of the system is expected to increase as
$(v/c)^2$ \citep{mcwilliams10,nlp+10}, where $v\sim(t_{\rm merge}-t)^{-1/8}$
traces the increasing orbital speed during inspiral, going to a
maximum of $v_{\rm max}\approx c/\sqrt{6}\approx 0.4c$ at the
innermost stable circular orbit, and with
a singularity at merger ($t_{\rm merge}$).  The flux
increase will be secular and may be detectable, but the timescales
over which it changes appreciably are likely either too long (during
the lengthy inspiral) or too short (right before merger), and will be
difficult to identify uniquely.

A promising candidate is variability induced by precession (also see
\citealt{katz97} for a related discussion). If there is a Poynting
flux associated with a jet, as in \citet{pll10}, the axis of this jet
could precess if the BH spins are not aligned with the orbital
angular momentum.  This would presumably cause the EM signal to vary
on that timescale (although it could be more complicated;
\citealt{katz97}).  The precession timescale is expected to be
$\tau_p\sim M(v/c)^{-5}$ \citep{acst94}.  The scale of the variations
is not known (it depends on the anisotropy of the emission), but could
easily be $>50$\%.

While a full treatment is beyond the scope of this paper, we are drawn
to consider the detectability of precessing jets in binary SMBHs for
two reasons.  First, the time spent at a moderate velocity $v/c\approx
0.1$ compared to the duration of the merger itself is large, scaling
as $(v/v_{\rm max})^{-8}$. A much larger number of systems exist in
this state compared to those merging; their timescales are more
amenable to detection.  Second, a system will undergo many precession
cycles, so periodic modulation may be detectable (along with other
changes, such as secular increase or orbital modulation); we expect
${\cal N}_p\sim (v/v_{\rm max})^{-3}$ periods to be visible in a
roughly logarithmic velocity range.  Precession has likely been seen
in galactic BH binaries \citep{katz97}, and does have observational
consequences for the jet emission.  In a future paper (O'Shaughnessy
et al., in prep), we will discuss the detectability of binary SMBH jet
precession in detail.

\acknowledgements We thank B.~Hughey, D.~Frail, and S.~Wyithe for
helpful comments.  DLK was partially supported by NSF award
AST-1008353.  ROS is supported by NSF award PHY-0970074.  MV
acknowledges support from SAO Award TM1-12007X and NASA awards ATP
NNX10AC84G and NNX07AH22G.


\end{document}